\documentclass[11pt]{amsart}

\usepackage{amsmath,amsfonts,amssymb}
\usepackage{dsfont}
\usepackage{array,delarray}
\usepackage[dvips]{graphicx}
\usepackage{xspace}
\usepackage{color}
\usepackage[square,sort&compress]{natbib}

\graphicspath{{./}{./figs/}}

\theoremstyle{definition}
\theoremstyle{remark}
\numberwithin{equation}{section}     

\long\def\Comment#1{
{
 \endIgnore{\relax} 
 }}
\def\endIgnore{\relax}

\begin{document}

\title
{Computer algebra in systems biology}

\author[R.~Laubenbacher]{Reinhard~Laubenbacher}
\address{Virginia Bioinformatics Institute and Mathematics Department\\
Virginia Polytechnic Institute and State University} 
\email[Reinhard~Laubenbacher]{reinhard@vbi.vt.edu}

\author[B.~Sturmfels]{Bernd~Sturmfels}
\address{Mathematics Department\\
University of California, Berkeley} 
\email[Bernd~Sturmfels]{bernd@math.berkeley.edu}

\date{\today}


\thanks{Reinhard Laubenbacher
acknowledges support by the National Science Foundation
(DMS-051144) and the National Institutes of Health (RO1
GM068947-01). Bernd Sturmfels was supported by
the National Science Foundation (DMS-0456960),
the Alexander von Humboldt Foundation, and the DARPA
Program {\em Fundamental Laws of Biology}.
We are grateful to Lior Pachter, Brandilyn Stigler, and
the referees for their comments.}

\maketitle

\section{Introduction}
Molecular biology has undergone a dramatic revolution during the
second half of the twentieth century, exemplified by the discovery
of the structure of DNA and the sequencing of the human
genome.  Since then a series of technological
advances has given experimentalists the ability to make ever-more
detailed measurements of an increasing number of molecular
components of the cell.  DNA microarrays, for instance, are small
silicon chips spotted with short segments of DNA that can be used
to measure the activity levels of thousands of different genes in
tissue sample extracts simultaneously.  Soon it might be possible to make
large-scale quantitative measurements in a single cell. Being able
to take such global snapshots of molecular processes has opened up
the possibility of studying the changes that are constantly going
on in cells as a coherent dynamical system with intricately
interacting parts, rather than studying the parts in isolation.
Thus, the new field of {\it systems biology} has emerged
\cite{Alon, kitano}.

Biological networks tend to be highly complex, with many variables
that interact with each other in nonlinear ways, making it
difficult to study such systems without the help of sophisticated
mathematical tools and concepts.  It is even unclear what the
right formal language should be for their description \cite{lazebnik}. 
A characteristic feature of systems
biology research is its heavy use of mathematical methods. One
tool which has been applied recently to biological problems is
{\em computer algebra}, a field of mathematics that combines the
ability of computers to carry out symbolic calculations with
concepts from abstract algebra. Computer algebra has been used in
the life sciences in a variety of ways, such as the construction
of phylogenetic trees encoding the evolutionary relationship
between different species \cite{cipra, marta}, or the
construction and analysis of models of intracellular
biochemical networks \cite{LS, yildirim}. For many more
such applications see \cite{Barnett, ASCB}.

This paper aims to introduce mathematicians to the new field of {\em algebraic
biology} by way of a very simple (simplified), well-studied biological network, which
can be explained and studied in this limited space. We
discuss two published  models for this network, one continuous
and one discrete, and analyze them using tools from
computer algebra. While the models are simple enough to be analyzed
by hand, the goal is to illustrate the applicability of the techniques.
And the reader can easily verify the software calculations. Along the
way we highlight mathematical challenges and research problems.

\section{Computer Algebra}\label{CompAlg}

Computer algebra provides tools for computing with symbols rather
than with floating point numbers. Software systems for computer
algebra include familiar commercial packages, such as {\tt Maple},
{\tt Mathematica}, or {\tt Magma}, as well as a wide range of more
specialized systems, many of which are free and often run faster
on specialized tasks. One important theme in computer algebra is
the solution of non-linear algebraic equations. In the context of
systems biology, this problem arises when one wishes to compute
the steady states of a dynamic model. As an example, consider the
following system of two equations where $x$ and $y$ are the
unknowns and $k_1$ and $k_2$ are parameters:
$$    x^2 \,+ \,k_1 xy - 1 \quad = \quad
     y^2 \,+\, k_2 xy - 1 \quad = \quad  0. $$
Using a computer algebra method known as {\it Gr\"obner bases} \cite{CLO, Notices,
ASCB},
the two given equations can easily be rewritten in the following equivalent form:
$$ (k_1 k_2-1) y^4 \,+ \,(k_2^2-k_1 k_2+2) y^2-1\, = \,
k_2 x + (1-k_1 k_2) y^3+(k_1 k_2 - k_2^2-1) y \,\,= \,\, 0 .$$ The
first equation involves no $x$. Using the quadratic formula, we
can therefore express $y$ in terms of the parameters $k_1,k_2$.
The second equation gives $x$ in terms of $y$ and $k_1,k_2$.
Further analysis reveals that there are always four real solutions
$(x,y)$ if  $k_1 k_2  < 1$ but only two real solutions if $k_1 k_2
> 1$.

For a second example, consider the following equations in 9 unknowns
which are  derived from the discrete model for the \emph{lac} operon 
in Section \ref{discrete_model}:
\begin{eqnarray*}
x_1&=&x_4x_5+x_4 \\
x_2&=&x_1 \\
x_3&=&x_1\\
x_4&=&1 \\
x_5&=&x_6x_7+x_6+x_7+1 \\
x_6&=&x_3x_8 \\
x_7&=&x_6+x_8+x_9+x_8x_9+x_6x_8+x_6x_9+x_6x_8x_9 \\
x_8&=&x_2\\
x_9&=&1
\end{eqnarray*}
We consider this as a system of equations over the Boolean field with two elements
$\{0,1\}$. The arithmetic in this field is characterized by the equation $1+1=0$.
A Gr\"obner basis for the given system consists of the expressions
$$ x_5, x_1+1, x_2+1,x_3+1, x_4+1, x_6+1, x_7+1, x_8+1, x_9+1.$$
(Note that this computation is carried out in the polynomial ring over the
field with two elements, so we need to account for the relations $x_i^2=x_i$,
when making this calculation.) From this Gr\"obner basis we can read off 
immediately that the only solution to 
this system is the 9-tuple $(1,1,1,1,0,1,1,1,1)$.

This answer could have been found without
Gr\"obner bases, by either solving the system ``by hand" or
by plugging in all $2^9$ binary vectors of
length $9$. However, the discrete dynamical systems that
are of interest in biology are now much more complex 
(due to advances in the experimental technologies, as argued above).
 For such systems, naive approaches will not work,
 and more sophisticated tools, such as computer algebra,
  are needed for the analysis.
For instance, one of the biochemical networks used in the recent DREAM
competition to reverse-engineer networks from data sets \cite{DR}
contains 58 molecular species, including mRNA, proteins, and metabolites. 
Data from this network can be used by
reverse-engineering methods, e.g., \cite{LS} to construct
a Boolean network model. This model would have $2^{58}$ states which precludes
an analysis by exhaustive enumeration of the state space. 
\section{The Lac Operon}

We illustrate the use of computer algebra in systems biology
by way of a gene regulatory network discovered by Jacob
and Monod \cite{JM}, earning them the 1965 Nobel Prize in Medicine.
 In prokaryotic organisms, some functions of gene 
regulation are accomplished by so-called \emph{operons}, groups of
genes that are adjacent to each other on the genome and are  transcribed
as a single segment of mRNA. Operons also have control
elements -- transcription factors -- that bind to regulatory
elements in the DNA and activate or inhibit the transcription of
structural genes. Transcription factors that stimulate
transcription are called inducers. They bind to regulatory
elements in DNA called promoters. Repressors, on the other hand,
bind to elements in DNA called operators and are effectively preventing transcription.

\begin{figure}[ht]
\centerline{ \raise1pt\hbox{ \framebox{
\includegraphics[width=.9\textwidth]{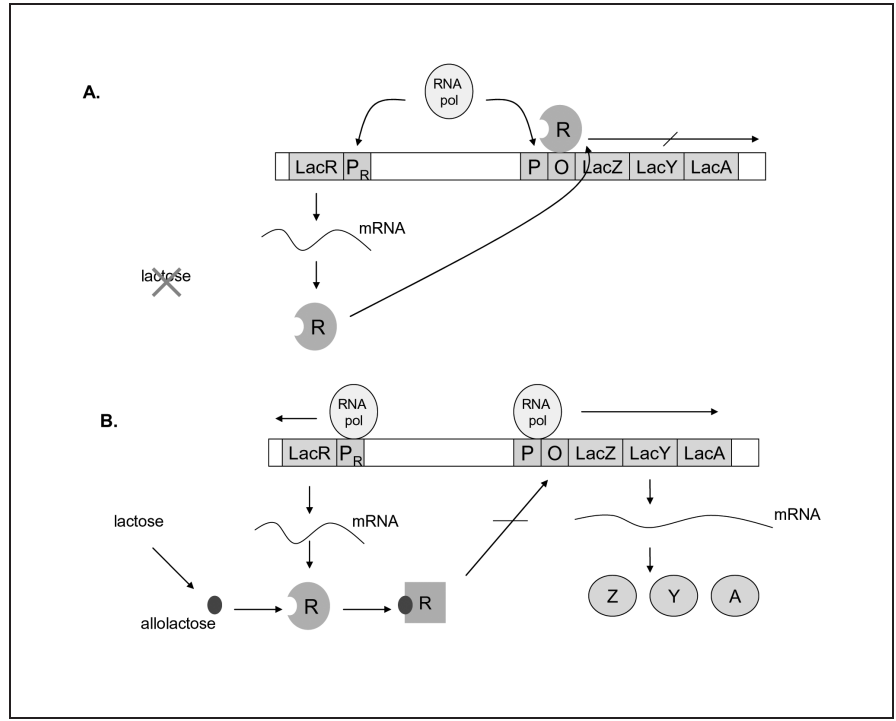}}}}
 \caption{The \emph{lac} operon. See \cite[p. 37]{deBoer}.}
\label{lac_operon}
\end{figure}

The \emph{lac} operon (Figure \ref{lac_operon}) in \emph{E. coli} is one such example. 
It enables \emph{E. coli} to metabolize lactose into glucose and other
products, in case glucose is not available directly. 
When cells grow in
glucose-based media,  the activity of the
enzymes involved in the metabolism of lactose is very low, even if
lactose is available. However, when glucose is exhausted from the
media and lactose is present, the
concentration of enzymes involved in lactose metabolism increases.
This  process is called {\em induction}~\cite{lodish}.

Figure \ref{lac_operon} is to be interpreted as follows. In panel A no lactose
is present. The repressor protein $R$ can bind to the operator
region of the \emph{lac} operon and block RNA polymerase from
transcribing the operon genes. In panel B lactose and its
isomer allolactose are present. The allolactose causes
a conformational change in the repressor protein which prevents
it from blocking transcription. Consequently, RNA polymerase transcribes
the operon into mRNA which, in turn, is translated into the proteins
$Z, Y,$ and $A$.

How is the \emph{lac} operon regulated? Positive control takes place in the 
absence of glucose and
presence of lactose in the cell. Some
of the lactose is converted into allolactose, which acts as an inducer
of transcription of the \emph{lac} genes. This process involves the protein CAP (catabolite activator protein), which forms a complex
with the substrate cAMP. This complex, in turn, binds to the activator region of the operon.
Translation of the
mRNA gives rise to the proteins $\beta$-galactoside permease (LacY),
which transports external lactose into the cell, and $\beta$-galactosidase (LacZ),
which cleaves lactose into glucose and its stereoisomer galactose. It also converts lactose into allolactose (Fig. \ref{lac_operon}.B). A third protein, LacA, is
not directly involved in lactose metabolism.  
Transcription continues until glucose becomes available. When
that happens two negative control mechanisms take over. 
Inside the cell, synthesis of the protein cAMP is inhibited, which is
needed for transcription of the operon, and the repressor protein LacI
can bind to the operator region of the lac operon, preventing transcription 
(Fig. \ref{lac_operon}.A).
This is known as \emph{catabolite repression}. Furthermore, external
glucose inhibits the transport of lactose into the cell through a process
called \emph{inducer exlusion}. We want to emphasize that this description 
is very simplified and does not mention many of the biochemical mechanisms
involved in this control circuit.

\section{A Continuous Model}
As one of the oldest known gene regulation network, the \emph{lac} operon has
been studied extensively and many different mathematical models
have been constructed for it; see \cite{leibler}.  The most common type
of model  is based on ordinary differential equations. For a model
from the recent literature see \cite{yildirim}. We discuss here
the very simple dynamical systems model described in Section 5.2
of the undergraduate text book \cite{deBoer}.  It consists of
three equations, modeling the concentration of the repressor $R$ and the rates
of change of the operon mRNA $M$ and allolactose $A$.  The three equations are:

\begin{eqnarray*}
R&=&\frac{1}{1+A^n},\\
\frac{dM}{dt}&=&c_0+c(1-R)- \gamma M ,\\
\frac{dA}{dt}&=&ML-\delta A-\frac{vMA}{h+A}.
\end{eqnarray*}
Here $c_0,c,\gamma,v,\delta,h$ and $L$ are certain model parameters,
$n$ is a fixed positive integer, and
the concentrations $R, M$ and $A$ are functions of time $t$.

This model is based on several assumptions.  For instance, we
do not distinguish between intracellular and extracellular
lactose, and denote both by $L$. Another assumption is that
$\beta$-galactosidase is proportional to operon activity $M$ and
is not represented explicitly.  The concentration of the repressor
$R$ is represented by a sigmoid function, a so-called {\em Hill
function}. When extracellular lactose is present and is
transported to the intracellular environment by lactose permease,
produced by the activity of the operon, the allolactose
concentration increases and inhibits the repressor $R$.  The
Hill coefficient $n$ determines the shape of the sigmoidal function.
The larger $n$, the steeper and more switch-like the curve becomes.
As we will see, this coefficient also determines the degree of the
polynomial system we need to solve. Biologically, $n$ can often be
interpreted as a specific measured quantity, such as the number of
molecules of a particular species that are involved in a cooperative
reaction.

The rate
of change of the gene transcripts $M$ is composed of a baseline
activity represented by the constant $c_0$, the concentration $A$
of allolactose (which inhibits the repressor $R$), and a
degradation term $\gamma M$.  The concentration of allolactose $A$
increases with the activity $M$ of the operon genes in conjunction
with the presence of lactose $L$. Its degradation (the terms on
the right-hand side with minus signs) is represented by a
Michaelis-Menten type enzyme substrate reaction composed of two
terms.  The parameters $c_0,c,\gamma,v,\delta,h$ and $L$ need to
be estimated using biological considerations or numerical methods,
to ensure that the model is consistent with experimental data.

This model is also quite simplified, both from a biological and a
mathematical point of view.  But even a simple model can be useful.
The purpose of modeling is to identify the essence of a system, that is, to identify the
components and dynamics that are key to conferring the biological function.
This is analogous to identifying that the key component for moving the bus forward is the
engine.
The art of constructing mathematical models of biological systems (or
any other type of system, for that matter) is to incorporate the
most important features and mechanisms and discard the
irrelevant ones. Comparing the model to the Boolean
network model constructed in the next section, we will see
certain basic similarities, even though the mathematics is different.
We have a time-discrete
finite dynamical system on the one hand and a continuous-time
system given by differential equations on the other hand.

It is now time to analyze the dynamics of the continuous model, by computing its steady
states.  We do this by phrasing the problem in a way that
 makes it amenable to using algebra,
namely  by setting the right hand sides of  the
differential equations to zero:
$$  c_0 \,+ \,c\cdot (1- \frac{1}{1+A^n}) \,-\,\gamma \cdot M
\,\,\,\, = \,\,\,\,
M \cdot L \,-\, \delta \cdot A \, - \,\frac{v M A}{h+A}
\,\,\,\, = \,\,\,\, 0 .$$
This is a system of two algebraic equations in two variables
$A$ and $M$, which depends on the various parameters.
Note that the steady state values for the missing variable
$R$ are determined  by the equation $\,R = 1/(1+A^n)$.

Following the discussion in \cite[Section 5.2]{deBoer}, we
leave the concentration $L$ of lactose unspecified
while the other parameter values are fixed as follows:
 $$ c =  \gamma =  v = 1 ,\,\,
c_0 = \frac{1}{20},\, h = 2,\, m = 5, \,
\delta = \frac{1}{5}.
$$
We also set $n=5$.  Our algebraic equations now take the form
$$
\frac{1}{20}\,+ \,\frac{A^5}{1+A^5} -M
\quad = \quad  M \cdot L - \frac{1}{5} \cdot A \,- \,\frac{MA}{2+A}
\quad = \quad 0 .$$
By clearing denominators and eliminating
the unknown $M$, we find that
$$ 4 \, A^7\,+ \,(29 -21L) \, A^6
\,- \,42 L \, A^5 \, + \,4 \, A^2 \,+ \,(9-L) \, A
\,- \,2 L  \,\,\, = \,\,\, 0 .
$$
This is a polynomial of degree $7$ in $A$. The
discriminant of this polynomial in $A$
is a complicated polynomial of degree $12$ in the parameter $L$.
This discriminant has precisely two positive roots, which we determine to be
$$  L_1 \,=\, 0.68454... \quad \hbox{and} \quad
L_2 \,=\, 1.51054....
$$
For all values of $L$ between $L_1$ and $L_2$,
there are three positive steady states. For example, if $L = 1$
then the steady states $(R,M,A)$
 of our system are
$$  (0.2272, 0.0506, 0.9994)\,,\,\,
(0.6907, 0.1859, 0.8642)\,,\,\,
(2.3717, 1.0368, 0.0132).
$$
The above expression $ \,4 A^7 + (29-21L) A^6 + \cdots \,$ is the equation of
the bifurcation diagram in the $(A,L)$-plane which is depicted in \cite[Figure 5.3(b)]{deBoer}.
It describes the steady-state allolactose concentration $A$ as a function
 of the lactose concentration $L$. As argued in \cite[Section 5.3]{deBoer},
 the emergence of these three steady states shows that this model correctly
 captures key features of the \emph{lac} operon. Computer algebra allows us to vary
 other parameters and enables us to conduct a very careful analysis of the dynamics of
 this model. In particular, using computer algebra,
 we can derive a precise algebraic description of the region in
 parameter space for which the dynamical system has more than one
 stationary point, and we can identify parameter values at which interesting
 phenomena (e.g.~Hopf bifurcations \cite{GMS}) might occur.


\section{A Discrete Model}\label{discrete_model}
While most models of the \emph{lac} operon have used the
framework of ordinary differential equations, other
modeling frameworks can also be useful. We describe a discrete model, 
in the form of a {\em Boolean network}, taken from
the recent article \cite{SV}. Like most models, it is very simplified in its
representation of biological details and mechanisms.  But it attempts
only to capture a basic dynamic feature of the
\emph{lac} operon, namely its bistability. 

Each one of the variables
in the model can take on the states 0 and 1, corresponding to the
simplifying biological assumption that the role a molecular species plays
in this network depends only on its absence or presence (defined by
a suitably chosen threshold) rather than a more refined measure of
its concentration in the cell. 
The model has nine variables, representing concentrations of various 
molecular species, some of which have already appeared in the continuous
model in the previous section:
\begin{enumerate}
\item
mRNA for the genes LacZ, LacY, and LacA ($M$) (the value of this
variable indicates whether the operon is ON or OFF), 
\item
\emph{lac} permease ($P$),
\item 
$\beta$-galactosidase ($B$),
\item
catabolite activator protein CAP ($C$),
\item
repressor protein LacI ($R$),
\item
lactose ($L$) and allolactose ($A$),
\item
low concentrations of lactose ($L_\ell$) and allolactose ($A_\ell$).
\end{enumerate}

The two variables $L_\ell$ and $A_\ell$ are ``artifical," in the sense that, for the
model to be accurate, it needs to allow for three, rather than two, possible concentration
levels of lactose and allolactose: absent, low, and high. In order to avoid the
use of multistate variables, we can keep the binary setting by introducing 
additional variables that account for low concentrations of these chemicals.
The model has two parameters: external lactose $a$ and external glucose
$g$. These are either present (1) or absent (0).
The interactions between these molecular species are described by
Boolean functions:

\begin{enumerate}
\item
$H_M(t+1) \, = \, \neg R(t)\wedge C(t)$, 
\item
$H_P(t+1) \, = \, M(t)$,
\item
$H_B(t+1) \, = \,M(t)$, 
\item
$H_C(t+1) \, = \, \neg g$,  
\item
$H_R(t+1) \, = \,\neg A(t)\wedge\neg A_\ell (t)$, 
\item
$H_A(t+1) \, = \, L(t)\wedge B(t)$, 
\item
$H_{A_\ell}(t+1) \, = \, A(t)\vee L(t)\vee L_\ell (t)$,
\item
$H_L(t+1) \, = \, \neg g\wedge P(t)\wedge a$,
\item
$H_{L_\ell}(t+1) \, = \, \neg g\wedge (L(t)\vee a)$.
\end{enumerate}

These functions are to be interpreted as follows. The genes are transcribed
($H_M(t+1)=1$) at time $t+1$ if the repressor $R$ is absent and CAP is present
at time $t$; permease and $\beta$-galactosidase are present at time $t+1$ if
the operon is transcribed at time $t$. The  interpretation of the other functions is similar.
 The result is a time-discrete dynamical system $\mathbf H(t)$,
on the space of binary 9-tuples, with dynamics generated by iteration of $\mathbf H$.
The model assumes that the molecular mechanisms leading
from activation of a gene to the production of the corresponding
protein (transcription plus translation) happen in one time step,
as does mRNA and protein degradation. 

It is shown in \cite{SV} that this simple model, based on very few
assumptions, displays a dynamic behavior that captures an essential
feature of the \emph{lac} operon, namely, its bistability.
To show this, one needs to examine the long term
dynamics, or steady states and periodic states,  of the model.
A state of the system is represented by a binary $9$-tuple 
$(M, P,B, C, R, A, A_\ell ,L, L_\ell)$, such as $(1,1,1,0,1,0,0,1,1)$.  
If we apply the dynamical system $\mathbf H$ to this state, with the
parameter setting $a=1, g=0$, that is, external lactose is present
and external glucose is absent, we obtain
the next state $(0,1,1,1,1,1,1,0,1)$. After iterating five more times
we reach the state $\mathbf u=(1,1,1,0,1,1,1,1)$, which turns out to be a steady
state, that is, $\mathbf H(\mathbf u)=\mathbf u$. This state corresponds to 
the operon being ON, with all molecular species, except the
repressor $R$, present. We would like to compute all such steady
states $\mathbf u$ for the system (1)-(9).

In the case of this model it is possible to solve the resulting system of
9 equations by hand, as mentioned earlier. Alternatively, one can find
the steady states
by computing the value of $\mathbf H$ on all $2^9$ 9-tuples for each of
the four possible parameter settings, for a total of 2048 evaluations
of $\mathbf H$. (For instance, the software 
\cite{DVD} can perform this computation.)
For larger models it is typically not possible to do either. For example, 
in \cite{LM} a Boolean network model of T-cell receptor signaling with 94
equations was published.
In that case, we can use computational tools 
provided by computer algebra, which provide a systematic way to solve the
system of steady state equations. The search for effective algorithms to 
solve very large systems of polynomial equations over finite fields, and
the Boolean field in particular, is currently an active area of research
(see, e.g., \cite{B}). We demonstrate their use with the
\emph{lac} operon model discussed here.

We first translate the Boolean functions in the model into polynomials.
This uses
arithmetic modulo 2.  To translate a Boolean function into a
polynomial function, we observe first that every Boolean function
is expressed using the logical operators $\wedge, \vee$, and
$\neg$.  These can be translated into polynomial operations by 
observing that the functions $a\wedge b$ and $a\cdot b$
take on the same Boolean values for given values of the variables.
That is, both functions take on the value 1 precisely if both $a$
and $b$ take on the value 1, otherwise the functions take on the
value 0.  Similarly, we see that $a\vee b = a + b +a\cdot b$ and
$\neg a = a + 1$.  If we apply this dictionary to the Boolean
functions in the model, while supressing the time variable,
 we obtain the following:
 \begin{eqnarray*} 
 H_M &=& (R+1)C=RC+C,\\ 
 H_P&=&M=H_B,\\
H_C&=& g+1,\\ 
H_R&=& (A+1)(A_\ell +1)=AA_\ell +A+A_\ell +1,\\
H_A&=&LB,\\
H_{A_\ell}&=&A+L+L_\ell +LL_\ell +AL+AL_\ell +ALL_\ell ,\\
H_L&=&(g+1)a P,\\ 
H_{L_\ell}&=&(g+1)(L+a L +a).
\end{eqnarray*}
A steady state of the system for a given choice of the parameters
$a$ and $g$ is one for which the functions do not
change the value of the variables.  It is therefore a
solution to the system of polynomial equations
$$
H_M \,=\, M, \,\, \ldots , \,  H_{L_\ell}\,=\,L_\ell .
$$
The system in Section \ref{CompAlg} arises for the parameter
choice $g=0$ and $a =1$. Solving the system for all four possible
parameter settings, $g=0=a; g=1, a = 0; g= 1 = a;$ and
$g=0,a =1$, results in the four corresponding steady states:
$$
\begin{matrix}
(0,0,0,1,1,0,0,0,0),&  (0,0,0,0,1,0,0,0,0), \\ \,\,(0,0,0,0,1,0,0,0,0), & (1,1,1,1,0,1,1,1,1).
\end{matrix}
$$
The last steady state is the one we found in Section \ref{CompAlg} and is the
only one for which the \emph{lac} operon is ON, in agreement with the biological
system. Note that, for each parameter setting, any initialization of the system will
end in the corresponding fixed point.

In \cite{SV} another smaller model is presented that captures some key features
of the feedback loop structure of this model and also shows the correct dynamics.
The main biological conclusion drawn in \cite{SV} from the models is that the dynamic behavior 
of the lac operon is due to properties of the network topology rather than 
the specific kinetics of the network that played a role in the continuous model discussed
in the previous section. 

\section{Discussion}
It is generally agreed that modern molecular biology can benefit
greatly from the use of mathematical techniques that allow the
construction of complex system-level models of biological
networks.  Conversely, the problems that arise in today's
biological research can provide important stimuli for mathematical
research.  This is aptly expressed in the title \emph{Mathematics
is Biology's Next Microscope, Only Better; Biology is Mathematics'
Next Physics, Only Better} of \cite{cohen}. We
have attempted here to describe through mathematical models of the
\emph{lac} operon how algebra can contribute to a formal
description and an analytical understanding of biological
phenomena.  One goal was to show that different types of
mathematical models (discrete and continuous) can provide insight
into biological mechanisms.  Furthermore, we have demonstrated
that computer algebra, not traditionally used in biology, is a
powerful tool that can help construct and analyze biological
models in a systematic way.  Thus, this paper should be viewed as an advertisement for
an in-depth study of the relationship between computer algebra, 
and mathematics in general, and systems biology.  The
marriage between the two promises to be extremely fruitful for
both.

One forum for such interactions is the annual international conference series
{\em Algebraic Biology} \cite{AB05} which was started in 2005. Another one
is the year-long program on {\em Algebraic Methods in Systems Biology and Statistics} 
held at the Statistical and Applied Mathematical Sciences Institute (SAMSI)
in North Carolina during the academic year 2008-09.

Several open mathematical research problems arise from our discussion. 
Firstly, in both Sections 3 and 4, we used
computer algebra algorithms to determine solutions to systems of nonlinear
polynomial equations. In one case the polynomials have real coefficients and in the other
they take binary values. In the binary case, a central problem is to improve
the power of the algorithms and the speed of the software implementations. 
Unlike the real coefficient case, relatively little work has been done until recently
on the development of efficient solution methods for polynomial systems
over finite fields in general, and the Boolean field in particular.  Where exact solutions cannot be computed,
information about the number of solutions, e.g., the number of steady states, would
be of interest. Tools involving zeta functions of algebraic
varieties over finite fields might be of help here. 
Another source of problems in computer algebra comes from the task of
inferring gene regulatory networks from experimental data sets, 
for instance, using the technique in \cite{LS}. 

There are numerous open problems concerning
continuous dynamical systems in systems biology.
Two problems we find particularly important 
are the characterization of chemical reaction networks that allow bistability  \cite{CFRS}
and the Global Attractor Conjecture for Toric Dynamical Systems~\cite{CDSS}.

\section{Biographical sketches}

{\bf Reinhard Laubenbacher} is Professor of Mathematics at Virginia Tech, a professor at the
Virginia Bioinformatics Institute, and Adjunct Professor in the Department of Cancer Biology
at the Wake Forest University School of Medicine. His research interests include computational
systems biology and cancer systems biology. With his
collaborators he has pioneered the use of tools and concepts from 
computational algebra to model biological systems.

\bigskip

{\bf Bernd Sturmfels} is Professor of Mathematics, Statistics and
Computer Science at UC Berkeley. A leading experimentalist
among mathematicians, he has authored ten books and about 180
articles, in the areas of combinatorics, algebraic geometry,
symbolic computation and their applications. Sturmfels currently
works on algebraic methods in statistics, optimization and
computational biology. His honors include the MAA's Lester
R. Ford Award and designation as a George Polya Lecturer.


\bibliographystyle{siam}

\end{document}